\documentclass[sigconf,table,screen=true]{acmart}
\AtBeginDocument{%
	\providecommand\BibTeX{{%
			\normalfont B\kern-0.5em{\scshape i\kern-0.25em b}\kern-0.8em\TeX}}}

\setcopyright{none}
\renewcommand\footnotetextcopyrightpermission[1]{}

\usepackage{xcolor,colortbl}
\usepackage{lipsum}

\usepackage[utf8]{inputenc}
\usepackage[T1]{fontenc}

\usepackage{amsmath,amssymb,amsfonts}
\usepackage[abbreviations]{foreign}
\usepackage{algorithm}
\usepackage{algpseudocode}

\usepackage{lstcustom}

\usepackage{graphicx}
\usepackage{setspace}
\usepackage{listings}
\usepackage{multirow}
\definecolor{backcolour}{rgb}{0.95,0.95,0.92}
\usepackage{subcaption}
\usepackage{textcomp}
\usepackage{booktabs}
\usepackage{xspace}
\usepackage{url}
\usepackage{pifont}
\usepackage[]{hyperref}
\usepackage{lipsum}

\usepackage[inline]{enumitem}
\setlist{noitemsep,topsep=0pt,parsep=0pt,partopsep=0pt}

\usepackage{setspace}
\usepackage[margin=5pt,font={stretch=0.9}]{caption}
\usepackage{fontawesome5}
\newboolean{showcomments}
\setboolean{showcomments}{true}
\ifthenelse{\boolean{showcomments}}
{ \newcommand{\mynote}[3]{
   \fbox{\bfseries\sffamily\scriptsize#1}
   {\small$\blacktriangleright$\textsf{\emph{\color{#3}{#2}}}$\blacktriangleleft$}}}
{ \newcommand{\mynote}[3]{}}

\definecolor{darkgreen}{rgb}{0.3,0.5,0.3}
\definecolor{darkblue}{rgb}{0.3,0.3,0.5}
\definecolor{darkred}{rgb}{0.5,0.3,0.3}

\newcommand{\sys}{\textsc{Montsalvat}\xspace}
\newcommand{\graal}{\textsc{GraalVM}\xspace}
\newcommand{\graalni}{\textsc{GraalVM} native-image\xspace}
\newcommand{\ecall}{\texttt{ecall}\xspace}
\newcommand{\ocall}{\texttt{ocall}\xspace}

\newcommand{\ocalls}{\texttt{ocall}s\xspace}

\makeatletter
\let\origsection\section
\renewcommand\section{\@ifstar{\starsection}{\nostarsection}}
\newcommand\nostarsection[1]{\sectionprelude\origsection{#1}\sectionpostlude}
\newcommand\starsection[1]{\sectionprelude\origsection*{#1}\sectionpostlude}
\newcommand\sectionprelude{\vspace{-7pt}}
\newcommand\sectionpostlude{\vspace{-2pt}}
\let\origsubsection\subsection
\renewcommand\subsection{\@ifstar{\starsubsection}{\nostarsubsection}}
\newcommand\nostarsubsection[1]{\subsectionprelude\origsubsection{#1}\subsectionpostlude}
\newcommand\starsubsection[1]{\subsectionprelude\origsubsection*{#1}\subsectionpostlude}
\newcommand\subsectionprelude{\vspace{-6pt}}
\newcommand\subsectionpostlude{\vspace{-2pt}}
\g@addto@macro\normalsize{%
  \setlength\abovedisplayskip{1pt}
  \setlength\belowdisplayskip{1pt}
  \setlength\abovedisplayshortskip{1pt}
  \setlength\belowdisplayshortskip{1pt}
  \setlength{\floatsep}{2pt}
  \setlength{\textfloatsep}{2pt}
  \setlength{\intextsep}{2pt}
  \setlength{\dblfloatsep}{2pt}
  \setlength{\dbltextfloatsep}{2pt}
}
\makeatother

\newcommand{\copyrighttext}{ \scriptsize \textcopyright 2021 ACM.               
	Personal use of this material is permitted.                                 
	Permission from ACM must be obtained for all other uses,                   
	in any current or future media, including reprinting/republishing this      
	material for advertising or promotional purposes, creating new collective   
	works, for resale or redistribution to servers or                           
	lists, or reuse of any copyrighted component of this work in other works.   
	This is the author’s version of the work. The final version is published in the proceedings of the 22nd International Middleware Conference. DOI: \href{https://doi.org/10.1145/3464298.3493406}{10.1145/3464298.3493406}}

\usepackage{tikz}
\usepackage{atbegshi}

\begin{document}

\title[Montsalvat: Partitioning Java Applications to Minimize the TCB in Intel SGX]{Montsalvat: Partitioning Java Applications to Minimize the TCB in Intel SGX}

\author{Peterson Yuhala}
\affiliation{%
	\institution{University of Neuchâtel}
	\city{Neuchâtel}
	\country{Switzerland}
}
\email{peterson.yuhala@unine.ch}

\author{J{\"a}mes Ménétrey}
\affiliation{%
	\institution{University of Neuchâtel}
	\city{Neuchâtel}
	\country{Switzerland}	
}
\email{james.menetrey@unine.ch}

\author{Pascal Felber}
\affiliation{%
	\institution{University of Neuchâtel}
	\city{Neuchâtel}
	\country{Switzerland}
}
	\email{pascal.felber@unine.ch}

\author{Valerio Schiavoni}
\affiliation{%
	\institution{University of Neuchâtel}
	\city{Neuchâtel}
	\country{Switzerland}	
}
\email{valerio.schiavoni@unine.ch}

\author{Alain Tchana}
\affiliation{%
	\institution{ENS}
	\city{Lyon}
	\country{France}	
}
\email{alain.tchana@ens-lyon.fr}

\author{Gaël Thomas}
\affiliation{%
	\institution{Télécom SudParis}
	\city{Institut Polytechnique de Paris}
	\country{France}	
}
\email{gael.thomas@telecom-sudparis.eu}

\author{Hugo Guiroux}
\affiliation{%
	\institution{Oracle Labs}
	\city{Z\"urich}
	\country{Switzerland}	
}
\email{hugo.guiroux@oracle.com}

\author{Jean-Pierre Lozi}
\affiliation{%
	\institution{Oracle Labs}
	\city{Z\"urich}
	\country{Switzerland}	
}
\email{jean-pierre.lozi@oracle.com}

\renewcommand{\shortauthors}{Yuhala, et al.}



\begin{abstract}
The popularity of the Java programming language has led to its wide adoption in cloud computing infrastructures.
However, Java applications running in untrusted clouds are vulnerable to various forms of privileged attacks.
The emergence of trusted execution environments (TEEs) such as Intel SGX mitigates this problem. 
TEEs protect code and data in secure \textit{enclaves} inaccessible to untrusted software, including the kernel and hypervisors. 
To efficiently use TEEs, developers must manually partition their applications into trusted and untrusted parts, in order to reduce the size of the trusted computing base (TCB) and minimise the risks of security vulnerabilities. 
However, partitioning applications poses two important challenges:
\emph{(i)}~ensuring efficient object communication between the partitioned components, and
\emph{(ii)}~ensuring the consistency of garbage collection between the parts, especially with memory-managed languages such as Java.
We present \sys, a tool which provides a practical and intuitive annotation-based partitioning approach for Java applications destined for secure enclaves. 
\sys provides an RMI-like mechanism to ensure inter-object communication, as well as consistent garbage collection across the partitioned components. 
We implement \sys with \graalni, a tool for compiling Java applications ahead-of-time into standalone native executables that do not require a JVM at runtime. 
Our extensive evaluation with micro- and macro-benchmarks shows our partitioning approach to boost performance in real-world applications up to $6.6\times$ (PalDB) and $2.2\times$ (GraphChi) as compared to solutions that naively include the entire applications in the enclave.
\end{abstract}


\newcommand{\copyrightnotice}{\begin{tikzpicture}[remember picture,overlay]       
	\node[anchor=south,yshift=2pt,fill=yellow!20] at (current page.south) {\fbox{\parbox{\dimexpr\textwidth-\fboxsep-\fboxrule\relax}{\copyrighttext}}};
	\end{tikzpicture}
}

\maketitle
\copyrightnotice
\pagestyle{plain}


\vspace{5pt}
\section{Introduction}
  
The Java programming language is widely used in cloud infrastructures.
Popular cloud frameworks such as Hadoop~\cite{hadoop}, ZooKeeper~\cite{zookeeper} and Spark~\cite{spark} are based on Java.
The recent growth of cloud-based services surrounding these popular tools raises security and privacy concerns. 
To address security issues in the cloud, major CPU vendors have introduced trusted execution environments (TEEs), such as Intel SGX~\cite{vcostan}, AMD SME~\cite{amd} and ARM TrustZone~\cite{trustzone}, which shield sensitive code and data inside secure memory regions called \emph{enclaves}.
In spite of their attractive security properties, programming TEEs is complex: it is usually done in compiled languages and low-level APIs at the function level, and it requires developers to make non-trivial efforts to minimise the trusted computing base (TCB).
Enforcing privacy with TEEs in a high-level, managed language like Java is particularly challenging.

Solutions exist to run entire applications (including the JVM) inside enclaves, while relying on a library OS~\cite{graphene,haven,sgxlkl,scone} to emulate unsupported OS logic in the enclave.
This approach offers good compatibility for legacy applications and requires little intervention from developers.
However, it significantly increases the size of the TCB: library OSs inside the enclave typically hit millions of lines of code~\cite{panoply}, which violates the principle of least privilege~\cite{saltzer1975protection} and increases the chances of enclave vulnerabilities.

Others (\eg, Civet~\cite{civet}, Uranus~\cite{uranus}) try to mitigate this problem by partitioning Java applications for enclaves.
Civet leverages static analysis~\cite{saBarros} to partition Java applications, but embeds a JVM and a library OS~\cite{graphene} inside the enclave, hence resulting in a large TCB. 
Uranus provides a technique to partition Java applications by annotating sensitive methods, but it requires developers to use third-party tools to infer trusted partitions of applications, as well as manual intervention by developers during the partitioning process.
One can manually partition specific Java frameworks for enclaves~\cite{vc3,secKeeper,opaque}, but this approach cannot be used for generic applications. 
Some systems like Glamdring~\cite{glamdring} propose techniques to partition native applications in C and C++ automatically, but they cannot be used for Java applications that rely on a managed runtime.

Partitioning Java applications for enclaves raises two significant challenges 
which were not sufficiently addressed by previous work:
\begin{enumerate}[leftmargin=*]
	\item Code running outside of an enclave (in the untrusted runtime) may allocate objects inside the enclave (in the trusted runtime), and code running inside the enclave may allocate objects outside of the enclave.
	Since both runtimes operate on separate memory heaps, there is a need for an efficient mechanism to ensure object communication across the two runtimes.
	
	\item Since there may be references between the untrusted runtime and the enclave, the garbage collector have to be extended to ensure consistency, \ie, objects in one runtime should not be destroyed if objects in the opposite runtime still reference them.
\end{enumerate}

To address these problems, we propose \sys, a tool that leverages annotations to partition Java applications into trusted and untrusted components automatically.
\sys leverages bytecode transformation to split Java applications between the trusted and untrusted runtimes, and applies distributed techniques like remote method invocation~\cite{rmi} to enable efficient communication between trusted and untrusted objects.
\sys introduces a dedicated GC helper to synchronise garbage collection (GC) between both runtimes.
Contrary to approaches based on library OSs~\cite{civet,graphene,sgxlkl}, \sys introduces a shim library in the enclave that relays unsupported libc calls in the enclave to the untrusted runtime, which reduces the TCB.

We implemented our approach with \graal~\cite{bonettagraalvm} and Intel SGX.
\graal is a high-performance JDK distribution that makes it possible to build and run applications implemented in a wide range of high-level languages (\eg, Java, Scala, JavaScript, Clojure, Kotlin, \etc).
\sys leverages a \graal tool named \emph{Native Image} supporting ahead-of-time (AoT) compilation of applications into native executables, called \emph{native images}~\cite{nativeImgs,fegade2020scalable,graalSite}, which do not require a JVM at runtime.
\graalni AoT compiles only reachable application methods, classes and fields, thereby excluding any redundant application logic from the final executable.
This results in quicker startup times and lower memory footprint for applications.
These properties of \graal native images make them particularly well-suited for restricted environments such as enclaves.
We evaluate \sys using synthetic benchmarks as well as real-world Java applications such as LinkedIn's PalDB~\cite{paldb} and GraphChi~\cite{graphchi}.
Our evaluation results show that partitioning PalDB and GraphChi can yield up to $6.6\times$ and $2.2\times$ performance improvements respectively, as compared to solutions that run these applications on a JVM in the enclave.


In summary, we propose the following contributions:

\begin{itemize}
	\item A practical and intuitive annotation-based approach for partitioning Java applications into insecure classes and secure classes to be run inside TEE enclaves.
	\item An RMI-like mechanism for inter-object communication across the trusted and untrusted runtimes.
	\item A garbage collection extension to synchronise object destruction across the trusted and untrusted heaps.
	\item Extensive experimental evaluation with various applications demonstrating the efficiency of our approach.
\end{itemize}

This paper is organised as follows. 
\S\ref{sec:background} provides background concepts and \S\ref{sec:related} discusses related work.
Our threat model is introduced in \S\ref{sec:tmodel}. 
The architecture and implementation of \sys are detailed in \S\ref{sec:architecture}. 
\S\ref{sec:evaluation} presents our extensive experimental evaluation, and we conclude in \S\ref{sec:conclusion}.

\vspace{5pt}
\section{Background}
\label{sec:background}
\vspace{5pt}
\subsection{Intel software guard extensions}

Intel software guard extensions (SGX) is an extension to the Intel instruction set architecture~\cite{sgx-shield} that enables applications to create \emph{enclaves}, \ie, secure isolated regions in memory. 
Enclave code and data are stored in a secure memory region, the enclave page cache (EPC). 
All EPC pages in DRAM are encrypted and only decrypted by a memory encryption engine (MEE) when they are loaded into a CPU cache line. 
Recent Intel processors support a maximum of 256\,MB of EPC memory (only 192\,MB are usable by SGX enclaves), limiting the amount of data in the enclave at any given time. 
The Linux SGX kernel driver can swap pages between the EPC and regular DRAM. 
This paging mechanism lets enclave applications use more than the total EPC, but at a significant cost~\cite{vault,secKeeper}. 

SGX applications are typically partitioned into \emph{trusted} and \emph{untrusted} parts that handle sensitive and non-sensitive operations, respectively.
Enclaves only run in user mode~\cite{vcostan}, hence OS services such as system calls cannot be executed directly inside of them and are instead relayed to the untrusted runtime. 
To enable communication across runtimes, Intel SGX provides \ecall and \ocall routines, which are specialised function calls that are used to respectively enter and exit an enclave. 
These calls induce costly context switches that last up to 13,100 CPU cycles~\cite{sgxperf,yuhala2021plinius}.


The Intel SDK~\cite{sgxsdk} makes it possible to partition and build enclave-based applications in C/C++ manually.
The SDK provides an enclave definition language (EDL) to define the enclave's interface, \ie, the set of all \ecall and \ocall routines.
An additional tool, \emph{Edger8r}, generates \emph{edge routines} using the EDL specifications.
The edge routines sanitise and marshal data into and out of the enclave.
All enclave code is then compiled into a final shared object that is cryptographically hashed for verification at runtime when it is loaded into enclave memory.

Manual partitioning can be avoided as solutions exist to easily run entire applications inside enclaves (\eg, SCONE~\cite{scone}, Graphene-SGX~\cite{graphene}, SGX-LKL~\cite{sgxlkl}, \etc).
However, these solutions have large TCBs, which degrade application performance and increase the chances of enclave vulnerabilities.

\subsection{\graalni}

\graalni is a tool, built on top of the \graal compiler~\cite{bonettagraalvm}, to compile ahead-of-time applications into standalone executables, which are named \emph{native images}.
It supports JVM-based languages, \eg, Java, Scala, Clojure and Kotlin.
Native images can also execute dynamic languages such as JavaScript, Ruby, R or Python~\cite{graalSite}.
\graalni leverages a points-to analysis~\cite{rountevPointsTo,nativeImgs,barnett2007annotations} approach to find all the reachable application methods that are compiled into the final native image, leading to faster startup times and lower memory footprint as compared to other Java runtimes.
\graalni enables applications to execute initialisation code (\eg, reading and parsing a configuration file) at build time, effectively reducing the application startup as less logic is executed at run time.
To transfer the result of the initialisation (Java objects) from build to runtime, \graalni takes a snapshot of the heap (called the \emph{image heap}) at the end of the build, and stores it into the generated executable.
The image heap is memory mapped inside the application heap at startup, allowing the application to start from the state initialised at build time.


\graalni makes a closed-world assumption, \ie, it considers that all application classes that can be executed at run time are known and available at build time. 
To support dynamic features such as reflection, the user provides a list of the classes, fields, and methods that can be accessed dynamically.
Each element of this list is then always included in the native image, in addition to all classes, fields and methods transitively reachable from these elements.
This list can be provided through e.g., CLI options, programmatically, or a JSON file.
\graalni provides a \emph{tracing agent}~\cite{agent} which assists developers in generating such a JSON file.




\graal native images do not run on a regular JVM (\eg, HotSpot):
runtime components that are needed to run JVM-based applications, such as a garbage collector, support for thread scheduling and synchronisation, as well as stack walking and exception handling are directly included inside the created native images. 

\looseness-1
\graalni provides the possibility of creating multiple independent VM instances at runtime, which are called \emph{isolates}.
Each isolate operates on a separate heap, allowing garbage collection to be performed independently.
Thus, threads executing in one isolate are not affected by garbage collection done in another isolate.
\sys creates a default isolate for each one of the two runtimes of the partitioned Java application (trusted and untrusted), which provides the execution contexts for all ``entry point'' methods (\eg, \texttt{main}). 



\section{Related work}
\label{sec:related}

We classify the related work into four categories:
\emph{(i)}~systems that make it possible to run full, unmodified applications inside enclaves,
\emph{(ii)}~framework-specific systems that support partial execution inside enclaves,
\emph{(iii)}~systems that allow for partitioning generic native applications, and
\emph{(iv)}~systems that allow for partitioning generic Java applications.

\looseness-1
\smallskip\noindent\textbf{Running full applications inside enclaves.}
Prior systems such as Haven~\cite{haven}, SCONE~\cite{scone}, Graphene-SGX~\cite{graphene} and SGX-LKL~\cite{sgxlkl} propose solutions to run entire legacy applications inside enclaves.
They introduce a library OS into the enclave to emulate OS logic.
For instance, SCONE leverages a modified version of the \textit{libc} to run microservices inside Docker~\cite{docker} containers.
While these solutions offer good compatibility with a wide range of applications and require low developer effort, they introduce millions of lines of code into the TCB.
This may significantly decrease enclave performance and leaves more room for security vulnerabilities.

\smallskip\noindent\textbf{Framework-specific partitioning.}
Some recent systems propose to manually partition specific frameworks and/or the applications that run on them into trusted and untrusted parts.
VC3~\cite{vc3} is a system for trustworthy data analytics in Hadoop that requires manually rewriting Map and Reduce functions to be used in SGX enclaves, while keeping the main Hadoop library outside the enclave.
SecureKeeper~\cite{secKeeper} proposes an extension to ZooKeeper which preserves confidential user data inside enclaves while maintaining the ZooKeeper framework outside the enclave.
The authors of Plinius~\cite{yuhala2021plinius} manually partition a persistent memory and machine learning (ML) library in order to enable efficient ML in SGX enclaves.
Opaque~\cite{opaque} is a secure data analytics platform built on top of Spark SQL that focuses on preventing access pattern leakage; it notably introduces SGX-enabled oblivious operators that can be used on tables that store sensitive data.
These systems help reduce the size of the TCB but focus on individual frameworks, which limits their flexibility.

\smallskip\noindent\textbf{Native code partitioning.}
Glamdring~\cite{glamdring} provides a technique to automatically partition C/C++ applications into untrusted and trusted parts using static program slicing.
Panoply~\cite{panoply} introduces \emph{micro-containers} which expose standard POSIX abstractions and run inside enclaves; applications must be refactored to extract sensitive code and data to be placed inside micro-containers.
These systems do not tackle the complexities introduced by managed languages.

\smallskip\noindent\textbf{Java code partitioning.}
Civet~\cite{civet} and Uranus~\cite{uranus} are two recent frameworks for running parts of Java applications inside enclaves.
Civet requires defining methods which will serve as the partitioning boundary, and Uranus requires annotating all sensitive methods. 
We argue that placing the enclave boundary at class level is more intuitive for developers. 
Additionally, our system benefits from \graal's optimisations, such as class initialisations at build time. 
Rather than a small shim library that relays unsupported calls to the trusted runtime as in our approach, Civet stores a full library OS inside the enclave (specifically, it uses Graphene SGX~\cite{graphene}), which leads to a larger trusted code base. 
CFHider~\cite{cfhider} also proposes to run parts of Java applications inside enclaves, but it specifically focuses on branch statement conditions with the objective to guarantee control flow confidentiality.




\vspace{5pt}
\section{Threat model}
\label{sec:tmodel}

\sys assumes enclave code and the CPU package are trusted, similar to related work with SGX~\cite{scone,glamdring,civet,uranus,vc3}. 
The source code annotation, image build via AoT compilation, and final enclave signing are done in a trusted environment. This prevents malicious classes/bytecode from being introduced into the enclave at runtime~\cite{r9049642}.
The integrity of the enclave can then be validated at runtime via remote  attestation~\cite{opera,vcostan} mechanisms.

\sys supports a powerful adversary with control over the full software stack, including the OS, hypervisors, and access to the physical hardware (\eg, DRAM, secondary storage, \etc).
The adversary's goal is to gain access to confidential data (\eg, passwords, encryption keys, \etc) which may be processed in trusted application classes, or to damage the integrity of confidential data. 

\sys is resilient to physical attacks like cold boot attacks~\cite{coldboot} aimed at reading sensitive data in DRAM, or bus probing~\cite{busprobing} to read the memory channel between the CPU and DRAM; the SGX security model~\cite{vcostan} prevents these.

We assume the adversary cannot physically open and manipulate the SGX-enabled processor package (as in~\cite{chen2021voltpillager}), and that the enclave code does not intentionally leak sensitive data.
Denial-of-service and side-channel attacks~\cite{van2018foreshadow,schwarz2017malware}, for which mitigations exist~\cite{oleksenko2018varys,gruss2017strong}, are considered out of scope.


\section{Architecture}
\label{sec:architecture}


\begin{figure}[!t]
	\centering
	\includegraphics[scale=0.65]{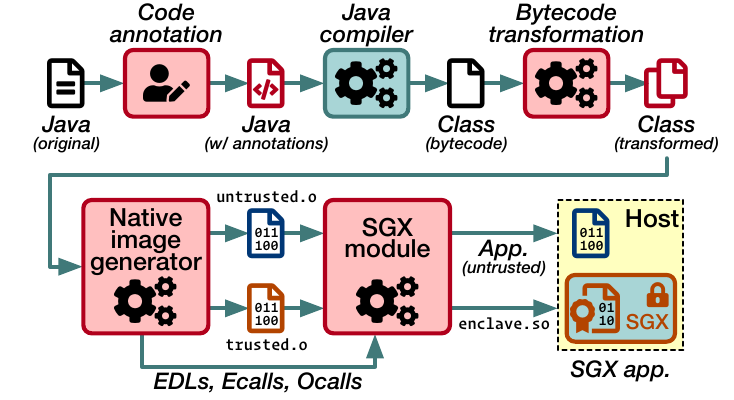}
	\caption{Overview of \sys's workflow. Annotated Java code, once compiled, is processed by the bytecode transfomer to produce two versions of the application classes. These classes are then used by the native image generator to produce trusted and untrusted images, which are finally linked with enclave libraries in the SGX module to build the final SGX application.}
	\label{fig:arch}
\end{figure}


The main goal of \sys is to partition Java applications for SGX enclaves. 
The final partitioned application includes a \emph{trusted} and an \emph{untrusted} part, respectively running inside and outside the enclave.
Figure~\ref{fig:arch} depicts \sys's complete workflow, from the source code to the generation of the final SGX application. 
It comprises 4 main phases:
\emph{(1)} code annotation,
\emph{(2)} bytecode transformation,
\emph{(3)} native image partitioning, and
\emph{(4)} SGX application creation. 

To illustrate the inner workings of \sys, we consider a synthetic Java application to be partitioned (see Listing~\ref{lst:main}).
Three classes mutually interact via method calls: classes \texttt{Account} and \texttt{AccountRegistry} perform sensitive operations, thus being secured in an enclave. 
However, class \texttt{Person} is untrusted and will not be included in the enclave.

\begin{figure}[t!]
\noindent\begin{minipage}[t]{.45\columnwidth}
\begin{lstlisting}[caption={},label={lst:main:1},frame=t]
@Trusted
public class Account {
	private String owner;
	private int balance;		
	public Account(String s, int b) {
		this.owner = s;
		this.balance = b;
	}	
	public void updateBalance(int v) {
		this.balance += v;
	}
}
\end{lstlisting}
\end{minipage}\hfill
\begin{minipage}[t]{.43\columnwidth}
\begin{lstlisting}[caption={},label={lst:main:2},firstnumber=13,xleftmargin=-12pt,frame=t]	
@Trusted
public class AccountRegistry {
	private List<Account> reg = 
		new ArrayList<Account>(); (*@\newline@*)						// Trusted in trusted obj
	public AccountRegistry() {}
	public void addAccount(Account a) {
		this.reg.add(a);
	}
}
\end{lstlisting}
\end{minipage}
\\[3mm]
\noindent\begin{minipage}[t]{.45\columnwidth}
\begin{lstlisting}[caption={},label={lst:main:3},firstnumber=22]
@Untrusted
public class Person {
	private String name;
	private Account account; (*@\newline@*)						// Trusted in untrusted obj
	public Person(String s, int v) {
		this.name = s;
		this.account = new Account(s, v);	
	}	
	public Account getAccount() {
		return this.account;
	}	
	public void transfer(Person p, int v) {
		p.getAccount().updateBalance(v);
		this.account.updateBalance(-v);
	}
}
\end{lstlisting}
\end{minipage}\hfill
\begin{minipage}[t]{.43\columnwidth}
\begin{lstlisting}[caption={},label={lst:main:4},firstnumber=38,xleftmargin=-12pt]
@Untrusted
public class Main {
	public static void main(String[] args) {
		Person p1 = new Person("Alice", 100);
		Person p2 = new Person("Bob", 25);
		p1.transfer(p2, 25);
		AccountRegistry reg =
		    new AccountRegistry();
		reg.addAccount(p1.getAccount());	
	}
}
\end{lstlisting}
\end{minipage}
\begin{lstlisting}[caption={Illustrative example with annotated classes using \sys: trusted classes (lines 1-12 and 13-21), and untrusted (lines 22-37 and 38-47).},label={lst:main},frame=b]
\end{lstlisting}
\end{figure}

\subsection{Partitioning language}
\label{sec:language}
When partitioning applications destined for enclaves, an important question to answer is how to specify what should be secured or not. 
Some recent work proposed annotation of sensitive data~\cite{glamdring}, others proposed annotation of sensitive routines~\cite{uranus,civet}. 
While these strategies work, we argue they are not always very intuitive for an application developer. 
Furthermore, they require complex and expensive data-flow analysis to ensure sensitive data is not leaked. 

Instead, we propose a technique based on \emph{class annotations}.
Classes are a fundamental building block for object-oriented applications, and it is very intuitive to reason about security along class boundaries.
Using annotations, developers can easily specify which classes need to be secured and which ones do not. Also, annotating whole classes instead of methods or fields prevents expensive data-flow analysis to track the propagation of sensitive data within a class in order to determine other sensitive methods or fields. Hence class annotation is a more pragmatic approach.

\sys supports two principal annotations: \texttt{@Trusted} and \texttt{@Untrusted}, which developers can use to specify secure and insecure classes, respectively.
In Listing~\ref{lst:main}, \texttt{Account} and \texttt{AccountRegistry} classes are annotated as trusted, whereas class \texttt{Person} is untrusted.

A trusted class will always be instantiated and manipulated inside the enclave, which has two main implications.
First, its member fields which are not instances of untrusted classes will be allocated on the enclave heap. 
Second, its methods are always executed inside the enclave.

Similarly, an untrusted class will have its instance objects allocated only on the untrusted heap, along with all its member fields which are not instances of trusted classes.
All its methods will be executed outside the enclave.

\sys maintains a single version of a trusted or an untrusted object in both worlds by leveraging \emph{proxy} objects (see \S\ref{sec:byte-code-partitioning}).
In our programming model, some classes can be \emph{neutral} as they may not be inherently trusted or untrusted.
This is the case for utility classes (\ie, \texttt{Arrays}, \texttt{Vector}, \texttt{String}) or other similar application-specific classes added by the developer.

\begin{lstlisting}[caption={Proxy for the untrusted class \texttt{Person}.},label={lst:person-proxy},frame=tb, float=t]
public class Person {
	private int hash;
	public Person(String s, int v) {
		byte[] buf = serialize(s);
		CCharPointer ptr = getPointer(buf);
		this.hash = getHash(this);
		ocall_relayPerson(this.hash, ptr, v);
	}
	public void Account getAccount() {
		ocall_relayGetAccount();
	}
	public void transferPerson(Person p, int v) {
		ocall_relayTransferPerson(p.getHash(), v);
	}
}	
\end{lstlisting}
Such classes are not security-sensitive and can be accessed in or out of the enclave without the use of proxies.
Contrary to trusted and untrusted classes, neutral class instances can have several copies in both worlds and may evolve independently.
The \texttt{@Neutral} annotation is optional, \ie, classes that are not annotated are by default neutral.

One may legitimately question the relevance of two annotations, thinking the \texttt{Trusted} annotation is sufficiently expressive.
Our argument for an \texttt{@Untrusted} annotation is twofold:
\emph{(1)} some classes may perform many system-related operations that are not supported inside enclaves, and keeping them in the enclave needlessly increases the TCB as they will perform many \ocall transitions to the outside;
\emph{(2)} classes which could introduce potential security vulnerabilities in the enclave should preferably be kept out of the enclave.
The \texttt{@Untrusted} annotation solves these problems, while also allowing for easy distinction with neutral classes.



\smallskip\noindent\textbf{Assumptions.}
We assume all \emph{annotated} classes are properly encapsulated (\ie, class fields are private). 
On the one hand, this prevents complex and expensive data flow analysis to ensure sensitive class fields do not leave the enclave. 
On the other hand, it guarantees that all class fields can only be accessed from outside classes via public getters and setters exposed by the class. 
As such, it is easier to control access to these sensitive class fields by applying techniques such as transparent encryption/decryption at the level of these public methods. 
Encapsulation being one pillar of object orientation, we believe this assumption to be reasonable.
\subsection{Bytecode transformation}
\label{sec:byte-code-partitioning}

The artefacts of the partitioned application consist of two native images: a \emph{trusted} and an \emph{untrusted} image. 
The trusted image will not have any untrusted functionality, and the untrusted image will not have any trusted functionality. 
However, trusted objects (\ie, instances of trusted classes) may call untrusted objects (\ie, instances of untrusted classes) and vice versa. 
Hence we need to have a bidirectional communication mechanism for code flow execution.
For that purpose, we introduce the notion of \emph{proxy classes}: instances of untrusted classes have proxies in the trusted runtime, and conversely instances of trusted classes have proxies in the untrusted runtime.
These proxies will serve as gateways to access the functionalities (\ie, methods) of their real class in the opposite runtime.

The proxy classes expose the same methods as the original classes and replace the method implementations by a transition logic to access the original functionalities across enclave boundaries. 
This design makes cross-enclave object communication easier and helps maintain the object-oriented nature of the program as a whole after it is partitioned.
We rely on bytecode transformations to \emph{create} these proxy classes and \emph{inject} code into existing classes to implement the enclave transitions.
\sys uses Javassist~\cite{jassist}, a popular bytecode transformation framework, to achieve this phase.

\sys automatically introduces matching proxy classes for all trusted and untrusted classes. 
The points-to analysis of \graalni automatically prunes/removes proxies for classes that are not reachable, which removes unnecessary proxies.
As \graal does not include unreachable proxy classes in the generated native images (see \S\ref{sec:native-image-partitioning}), we did not include that analysis in the bytecode transformer.
Listings~\ref{lst:person-proxy}, \ref{lst:account-proxy} and \ref{lst:account-relay} illustrate the result of bytecode transformations for the corresponding classes.


\begin{lstlisting}[caption={Proxy for the trusted class \texttt{Account}.},label={lst:account-proxy},frame=tb,float=t]
public class Account {
	private int hash;
	public Account(String s, int b) {
		byte[] buf = serialize(s);
		CCharPointer ptr = getPointer(buf);
		this.hash = getHash(this);
		ecall_relayAccount(this.hash, ptr, b);
	}
	public void updateBalance(int v) {
		ecall_relayUpdateBalance(this.hash, v);
	}
}	
\end{lstlisting}

For the purposes of the trusted image, this process creates proxy classes for untrusted classes by stripping the methods (\ie, removing the method bodies) of the untrusted classes.
Listing~\ref{lst:person-proxy} shows the corresponding proxy class for the untrusted class \texttt{Person} in our illustrative example.
The bodies of the stripped methods are replaced with native routines which will perform \ocall transitions to the corresponding method in the untrusted runtime (lines 7, 10, 13 in Listing~\ref{lst:person-proxy}).
Analogously, for untrusted image generation, the bytecode transformer creates proxy classes for trusted classes by stripping all methods of trusted classes, and replacing the method bodies with native methods which will perform \ecall transitions.
Listing~\ref{lst:account-proxy} shows the corresponding proxy class for the trusted class \texttt{Account} in our illustrative example.
The proxy class fields are removed and a \texttt{hash} field is added to each proxy class which stores the hash of the proxy object (\ie, line 6 in Listing~\ref{lst:account-proxy}). Our present implementation uses a hash function based on Java identity hash codes. To minimize hash collisions, a hashing algorithm like MD5~\cite{rivest1992md5} should be used.
The result of the stripping operations is the removal of all untrusted functionality from the trusted runtime, and conversely.
Only annotated classes are modified by the bytecode weaver, \ie, neutral classes are not changed.

In the rest of the paper, we refer to all unstripped classes as \emph{concrete classes} and stripped classes as \emph{proxy classes}.
We respectively call the instances of these classes \emph{concrete objects} and \emph{proxy objects}.
If a concrete object in one runtime (trusted or untrusted) has a correspondence with a proxy object in the opposite runtime, we refer to that concrete object as a \emph{mirror object} (\ie, the proxy's mirror copy).


\smallskip\noindent\textbf{Relay methods.}
For the methods in one runtime (a native image) to be callable from the other runtime (another native image), these methods must be exported as \emph{entry points}.
\graalni provides an annotation (\texttt{@CEntryPoint}~\cite{centrypoint}) for specifying entry point methods which can be callable from C.
These entry point methods must be static, they may only have non-object parameters and return types, \ie, primitive types or \texttt{Word} types (including pointers)~\cite{graalSite}, and they must specify the \graal isolate that will serve as execution context for the method.
As a result of these restrictions, it is not feasible to export all methods of concrete classes as entry points directly, as this would require changing their signatures.
To circumvent this limitation, we introduce static entry point methods to act as wrappers for the invocations of the associated class or instance methods.
We call these \emph{relay methods}. 

For every public method in a concrete class, including all constructors, the bytecode transformer adds an associated relay method to the class.
The native methods (\ie, \ecall and \ocall routines) we added to the stripped methods of the proxy classes will perform enclave transitions to invoke the corresponding relay methods.

\begin{lstlisting}[caption={Concrete class \texttt{Account}, once transformed.},label={lst:account-relay},frame=tb,float=t]
public class Account {
	private String owner;
	private int balance;		
	public Account(String s, int b) {...}
	public void updateBalance(int v) {...}

	@CEntryPoint
	public static void relayAccount(Isolate ctx, int hash, CCharPointer buf, int b) {
		String s = deserialize(buf);
		Account mirror = new Account(s, b);
		mirrorProxyRegistry.add(hash, mirror);
	}
	@CEntryPoint
	public static void relayUpdateBalance(Isolate ctx, int hash, int v) {
		Account mirror = mirrorProxyRegistry.get(hash);
		mirror.updateBalance(v);
	}
}
\end{lstlisting}

The parameters of a relay method comprise: an isolate which provides the execution context for the method call, the hash of the calling proxy object (for non-static methods), all primitive parameters of the associated method, and pointers (\ie, \texttt{CCharPointer}~\cite{ccharpointer}) which represent the addresses of buffers obtained from the serialization of any object parameters which are instances of neutral classes (as these classes do not need proxies).
For proxy object parameters, the hash of the corresponding proxy is passed as parameter and the corresponding mirror object will be used as the parameter once the real (\ie, concrete) method is called in the opposite runtime.
Similarly, for mirror object parameters, the hash of the corresponding proxy is also sent, and the corresponding proxy object is used in the opposite runtime as parameter in the method.

As for the serialized buffers, they are deserialized and the corresponding object parameter recreated in the body of the relay method.
For relay methods of constructors, we add code to instantiate the corresponding mirror object, as well as code to add the mirror object strong reference and associated proxy hash to a global registry, which we call the \emph{mirror-proxy registry}.
For instance methods, we add code to look up the corresponding mirror object in the registry, and then invoke the instance method on that mirror object with its corresponding parameters.
Neutral object return types from the untrusted runtime are also serialized and copied across the enclave boundary.
Both the trusted and untrusted runtimes have a mirror-proxy registry.

The code in Listing~\ref{lst:account-relay} outlines the state of concrete class \texttt{Account} after bytecode transformation.
For illustration, the relay method \texttt{relayAccount} (line 8) is added into concrete class \texttt{Account} in the trusted runtime automatically.

We complete the transformation of the other classes of our illustrative example as follows.
For proxy class \texttt{Person}, we will have \ocalls instead to the relay methods.
For proxy class \texttt{AccountRegistry}, we have a proxy \texttt{Account} as parameter in the \texttt{addAccount} method, and only its hash will be passed to the opposite runtime.
The resulting proxy method and the corresponding relay method in concrete class \texttt{AccountRegistry} are shown in Listing~\ref{lst:proxyrelay}.

\begin{lstlisting}[caption={Proxy (top) and relay (bottom) methods for the \texttt{AccountRegistry} class.},label={lst:proxyrelay},frame=tb,float=t]
public void addAccount(Account acc) {
	ecall_relayAddAccount(acc.getHash());
}

@CEntryPoint
public static void relayAddAccount(Isolate ctx, int hash) {
	Account mirror = mirrorProxyRegistry.get(hash);
	this.addAccount(mirror);
}
\end{lstlisting}


\smallskip\noindent\textbf{Native transition methods.}
The native transition methods (\eg, \texttt{ecall\_addAccount}) are C routines which perform enclave transitions to the opposite runtime.
At the time of bytecode transformation, the definitions of the native transition methods are absent and only their signatures are provided.
Their definitions will be generated by the native image generator (see the next section).



\subsection{Native image partitioning}
\label{sec:native-image-partitioning}

The \graal native image generator is responsible for building native images.
It takes as input compiled application classes (bytecode) and all their associated external libraries, including the JDK. 
The native image generator then performs points-to analysis~\cite{nativeImgs} to find the reachable program element (classes, methods and fields). 
Only reachable methods are then compiled ahead-of-time into the final native image. 
For an SGX-based environment, this let us exclude any redundant application logic from the enclave. 
The resulting image embeds runtime components for garbage collection (memory management), thread scheduling, \etc. 

\vspace{-2pt}
\smallskip\noindent\textbf{Static analysis for trusted and untrusted image generation.}
The bytecode transformations produced two sets of class files: the first set $(T)$ comprises modified trusted classes and untrusted proxy classes, while the second set $(U)$ comprises modified untrusted classes and secure proxy classes. 
The unannotated/neutral classes $(N)$ were not changed by the bytecode transformer. 
These three sets of classes are used by \graal to generate two native images, \ie the partitioned application.

The native image generator in \sys uses set $(T\cup N)$ as input for \emph{trusted image} generation and the set $(U\cup N)$ as input for \emph{untrusted image} generation.
To determine reachable program elements (\ie, classes, fields and methods) the native image generator leverages a static analysis technique known as points-to analysis~\cite{nativeImgs,balatsouras}.
Points-to analysis starts with all entry points and iteratively processes all transitively reachable classes, fields and methods~\cite{nativeImgs}.
For the sake of brevity, we do not include all the steps performed during points-to analysis in \graalni (see~\cite{nativeImgs} for details).
For the trusted image, all the relay methods of trusted classes will serve as entry points (recall the \texttt{@CEntryPoint} annotation, \eg, in Listing~\ref{lst:account-relay}).
For the untrusted image, the main entry point (Java application's \texttt{main} method) and the relay methods of untrusted classes will serve as entry points.
Conceptually, we can include the main entry point in either the trusted or untrusted image.
However we chose to add it in the untrusted image because:
\emph{(1)} it prevents an \ecall transition to invoke the main method and \ocall transition to create garbage collection helper threads (see \S\ref{sec:gc}) once in the main method, and
\emph{(2)} it is in accordance with Intel SGX's programming convention, as all SGX applications begin in the untrusted runtime.

\begin{figure}[!t]
	\centering
	\includegraphics[scale=0.5]{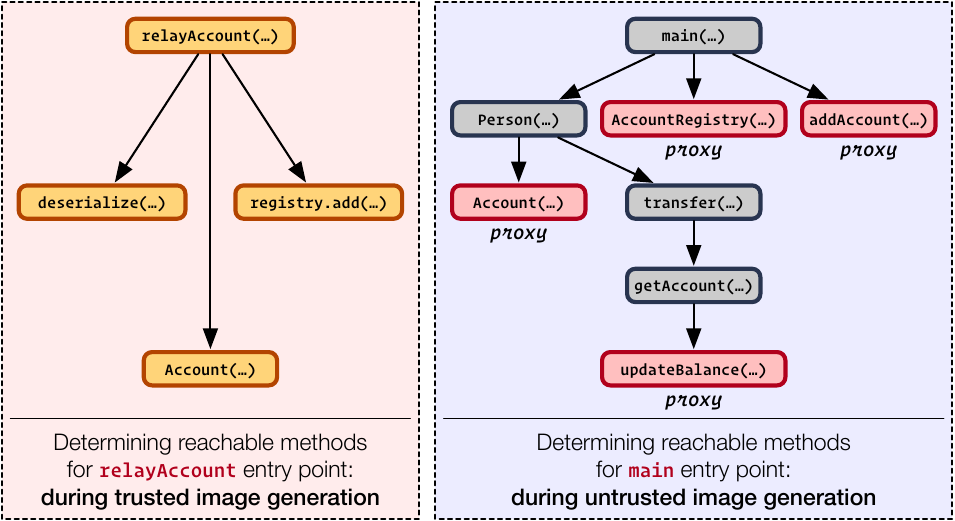}
	\caption{For the trusted image, the relay methods in the trusted classes ensure that other trusted class methods, as well as methods from neutral classes (\eg, \texttt{serialize}, \texttt{registry.add}, \etc), are reachable.
	Similarly, for the untrusted image, the main entry point ensures that the \texttt{Person} class methods, as well as methods from proxy classes (\texttt{Account} and \texttt{AccountRegistry}), are reachable.
	This is a subset of the full graph (other methods will be made reachable at leaf nodes).}
	\label{fig:call-graphs}
\end{figure}

Figure~\ref{fig:call-graphs} illustrates a simplified reachability analysis done for two entry point methods (\texttt{relayAccount} and \texttt{main}) to determine their reachable methods. 
A similar process is performed for all other entry points.

Once static analysis is complete, the trusted image no longer contains untrusted methods/functionality. 
It embeds proxies instead, in case some untrusted proxy class methods were reachable. 
Similarly, the untrusted image does not contain trusted methods, but only proxies to those if some proxy class methods were reachable. 
Following from our illustrative example, proxy class \texttt{Person} will not be included inside the trusted image since it is not reachable from any of the trusted classes.

In the \texttt{main} method in our illustrative example, at runtime during object creation, the constructors of \texttt{Person} $p1$ and $p2$ will instantiate a proxy object of the \emph{proxy} \texttt{Account} class. 
The string parameters \texttt{``Alice''} and \texttt{``Bob''} will be serialized and an \ecall transition will be made to create a corresponding \emph{mirror} \texttt{Account} object in the enclave. 
Similarly, when \texttt{p1.transfer(p2, 25)} is called, the corresponding proxy \texttt{Account} objects will perform enclave transitions to update the balances in the enclave. 
In the same way, the call to the proxy \texttt{AccountRegistry} constructor performs a transition to create a mirror object in the enclave corresponding to proxy object \texttt{reg}. 
The latter performs a transition too when \texttt{addAccount} is called.

By default, the native image generator compiles all reachable methods and links the latter with \graal's native libraries to produce an executable or a shared object file. 
We modified the image generator to bypass the linking phase that produces executables or shared objects, so as to produce relocatable object files (\texttt{.o}) which can be linked to other libraries to build the final SGX application. 
The resulting images for the trusted and untrusted parts are \texttt{trusted.o} and \texttt{untrusted.o} respectively. 
These will be dispatched to the SGX module and used to build the final SGX application.


\smallskip\noindent\textbf{SGX code generator.}
During bytecode transformation, native \ecall and \ocall transition routines are added to proxy classes.
We extended \graalni with a class to generate C code definitions for the corresponding \ecall (added in trusted proxy classes) or \ocall (added in untrusted proxy classes) transitions, as well as their corresponding header files.
Listing~\ref{lst:ecall-addAccount} shows the generated C code for the \texttt{ecall\_relayAddAccount} method.

\begin{lstlisting}[caption={C code for \texttt{ecall\_relayAddAccount}.},label={lst:ecall-addAccount},frame=tb,float=t]
void ecall_relayAddAccount(int hash) {
	Isolate ctx = getIsolate(); (*@\hfill@*)// Get the enclave isolate
	relayAddAccount(ctx, hash);
}
\end{lstlisting}


The code generator also creates associated EDL files used by the \emph{Edger8r} tool in the Intel SGX SDK to build bridge routines, which marshal the data across the enclave boundary.
The generated files are dispatched to the SGX module.

\subsection{SGX application creation}
\label{sec:sgx-app-creation}

This is the final stage in the \sys workflow.
Because SGX enclaves operate only in user mode, they cannot issue system calls and standard OS abstractions (\eg, file systems, network), which are ubiquitous in real-world applications. 
The solution to this problem is to relay these unsupported calls to the untrusted runtime, which does not have the same limitations. 
In contrast to systems that introduce a \emph{LibOS} (\ie, an entire operating system implemented as a library) in the enclave, we leverage an approach which involves redefining unsupported \texttt{libc} routines as wrappers for \ocalls. 
These redefined \texttt{libc} routines in the enclave constitute \sys's \emph{shim library}. 
The latter intercepts calls to unsupported \texttt{libc} routines and relays them to the untrusted runtime. 
A \emph{shim helper library} in the untrusted runtime then invokes the real \texttt{libc} routines. 
This by design reduces the TCB when compared to \emph{LibOS}-based systems.

\sys then compiles all generated \ecall routines and statically links them with the trusted image (\texttt{trusted.o}), the shim library and native libraries from \graal to produce the final enclave shared library, which corresponds to the trusted part of the Java SGX application. 
Similarly, the generated \ocall routines are also compiled and linked with the untrusted image (\texttt{untrusted.o}) and \graal native libraries to produce the final untrusted component. 
In accordance with Intel SGX's application model, \sys compiles the main entry point of partitioned applications in the untrusted component. 
The resulting trusted and untrusted components compose the final SGX application. 
At runtime, a \graal isolate is created in both the trusted and untrusted part of the application. 
These isolates provide the execution contexts for transition routines, \ie, the trusted isolate serves \ecall routines while the untrusted isolate serves \ocall routines.



\subsection{Garbage collection}
\label{sec:gc}

Following our partitioned application design, untrusted code objects can have trusted counterparts (proxies) and vice versa. 
This presents a challenge at the level of garbage collection (GC) because we must ensure synchronised destruction of objects across the trusted and untrusted heaps. 
More specifically, we need to synchronise GC of proxy and mirror objects, \eg, the mirror of proxy object \texttt{reg} should not be destroyed before \texttt{reg} in our illustrative example (Listing~\ref{lst:main}). 
Similarly, when \texttt{reg} is destroyed, its corresponding mirror object should be made eligible for GC.
 
Java provides \textit{finalizer methods}~\cite{javaSpecs} which the garbage collector invokes prior to garbage collecting an object. So one could envision a solution based on finalizer methods. However, the latter are deprecated since Java 9 and have badly designed semantics~\cite{graalFinalizers}. For example, a finalizer method can make a proxy object reachable again, which will lead to an inconsistent state across the trusted and untrusted heaps after the proxy’s mirror object is destroyed.

To address this problem, we implemented an application-level GC helper based on \texttt{weak references}. 
When a proxy object is created, \sys stores a weak reference and the hash of the former in a global list.
We use a weak reference here because it does not prevent the proxy object from being garbage collected once it is eligible for GC. 
The GC helper thread periodically (\eg, every second) scans this list for null referents of weak references, \ie, objects referred to by the weak references. 
Once such a null referent is found, it means the proxy object has been (or is eligible for being) garbage collected, and thus we can remove the corresponding mirror object from the mirror-proxy registry in the opposite runtime.
This makes the mirror object eligible for GC if it is not strongly referenced anywhere else.
Both the trusted and untrusted runtimes maintain independent lists of proxy weak references for the associated runtime, and two GC helper threads are spawned in the application: one to scan the trusted list in the enclave, and the other to scan the untrusted list.


%
%
%
%
%

\subsection{Running unpartitioned native images}

Despite the benefits of partitioning an enclave application, situations may arise where it is much easier for the application developer to run the entire application as a native image inside the enclave.
This could happen when the majority of classes potentially deal with sensitive information and no classes qualify as untrusted.
Consequently, \sys makes it possible to run unpartitioned applications. 
Unpartitioned applications do not require annotations, hence no bytecode modifications are performed. 
The original application is built into a single native image which is linked entirely to the final enclave object.

\subsection{Prototype implementation}
\label{sec:implementation}

Our current \sys prototype is based on \graal CE v21.0.0 for Java 8.
\graal and the bytecode transformer are implemented in Java.
Our modifications in \graal amount to $\sim$1,400 lines of code (LOC).
The bytecode transformer relies on Javassist v3.26 and contains $\sim$2,100 LOC.
The SGX module is based on SGX SDK and SGX driver v2.11.
It consists of $\sim$10,200 C/C++ LOC.
We plan to release \sys as open-source.

\section{Evaluation}
\label{sec:evaluation}


This section presents an experimental evaluation of \sys based on micro- and macro-benchmarks with real-world applications.
We seek to answer the following questions:

\begin{itemize}[]
	\item[$Q_1$)] What is the cost of proxy object creation and remote method invocations? (\S\ref{eval-proxy-creation}, \S\ref{eval-rmi-serialization})
	\item[$Q_2$)] How does partitioning impact GC performance? (\S\ref{eval-gc})
	\item[$Q_3$)] How does partitioning impact application performance? (\S\ref{eval-perf-partitioning})
	\item[$Q_4$)] How do partitioned and unpartitioned native images in SGX enclaves compare with applications running on a JVM in enclaves? (\S\ref{eval-comparison-jvm})
\end{itemize}

\subsection{Experimental setup}

Our evaluation is conducted on a server equipped with a quad-core Intel Xeon E3-1270 CPU clocked at 3.80\,GHz, and 64\,GB of DRAM. 
The processor has 32\,KB L1i and L1d caches, 256\,KB L2 cache and 8\,MB L3 cache. 
The server runs Ubuntu 18.04.1 LTS 64\,bit and Linux kernel 4.15.0-142. 
We run the Intel SGX platform software, SDK and driver version v2.11.
The EPC size on this server is 128\,MB, of which 93.5\,MB is usable by enclaves.
The enclaves have maximum heap sizes of 4\,GB and stack sizes of 8\,MB.
All native images are built with a maximum heap size of 2\,GB.

We use SCONE to run unmodified applications on a JVM in SGX enclaves. 
The SCONE containers are based on Alpine Linux~\cite{alpine}. 
The base SCONE image ships OpenJDK8 (tag: \texttt{8u181-jdk-alpine-scone5.1.0}).
For non-SCONE experiments, we execute the \graal compiler, which generates the native images, in OpenJDK-8u282. At execution, only the code of the generated native images runs.
All reported latencies are averaged over 5 runs.

\subsection{Performance of proxy creation}
\label{eval-proxy-creation}

\begin{figure}[!t]
	\centering
	\includegraphics[scale=0.67]{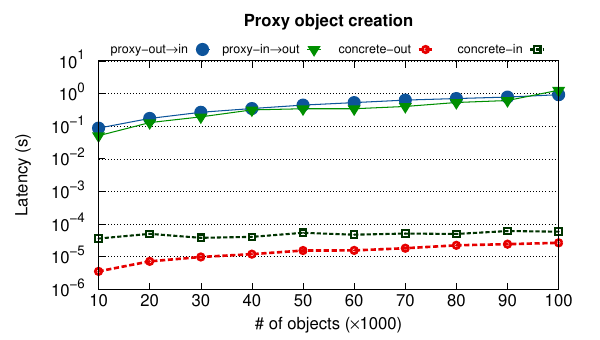}
	\caption{Performance of proxy object creation vs. concrete object creation.}
	\label{fig:proxy-creation}
\end{figure}

\emph{(Answer to $Q_1$)}
The goal of this experiment is to study the latency of proxy creation in relation to normal (concrete) object creation.
The notion of \emph{proxy} is an internal feature of \sys which could impact application performance.
We use a synthetic Java program to realise this experiment.
Figure~\ref{fig:proxy-creation} shows the results obtained.
We perform object instantiations in four different scenarios: concrete object creation in and out of the enclave methods (respectively labelled \texttt{concrete-in} and \texttt{concrete-out} in Figure~\ref{fig:proxy-creation}), and proxy object in (\texttt{proxy-in$\to$out}) and out (\texttt{proxy-out$\to$in}) of the enclave.
Scenario \texttt{concrete-out} corresponds to the base line for this experiment.

We observe that proxy object creation latency in the enclave is 3 orders of magnitude higher when compared to concrete object creation in the enclave, and proxy object creation latency out of the enclave is 4 orders of magnitude higher when compared to concrete object creation out of the enclave.
This performance drop when creating proxy objects is mainly due to the expensive enclave transitions required to instantiate the corresponding mirror objects in the opposite runtime.


\subsection{Performance of RMI and impact of serialization}
\label{eval-rmi-serialization}

\begin{figure}[!t]
	\centering
	\includegraphics[scale=0.67]{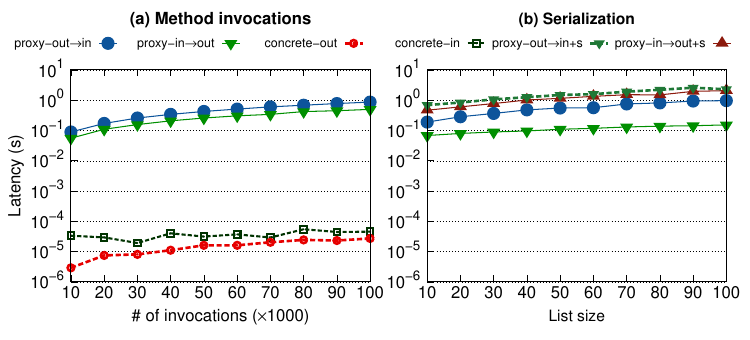}
	\caption{Performance of remote method invocations (RMIs) by proxy objects vs. concrete object invocations, and impact of serialization on RMIs.}
	\label{fig:rmi}
\end{figure}

\emph{(Answer to $Q_1$)}
The goal of this experiment is to study the performance of remote method invocations by proxy objects, and understand the impact of serialization on these invocations.
To this end, we generate synthetic programs where objects perform method invocations in four different scenarios: concrete object invoking instance methods in and out of the enclave (respectively labelled \texttt{concrete-in} and \texttt{concrete-out} in Figure~\ref{fig:rmi}\,(a)) and proxy object invoking instance methods remotely from within (\texttt{proxy-in$\to$out}) and out of (\texttt{proxy-out$\to$in}) the enclave without serializable parameters.
We vary the number of method invocations of the objects in these scenarios and calculate the corresponding latency of the invocations. 

Figure \ref{fig:rmi}\,(a) shows that when there is no serialization involved, the latency of proxy object RMI in the enclave is 3 orders of magnitude higher than the latency of concrete object method invocation in the enclave.
On the other hand, the latency of proxy object RMI out of the enclave is 4 orders of magnitude higher when compared to concrete object method invocation latency.
This overhead is similar to that observed in proxy object creation, and is mainly due to the expensive enclave transitions involved. 

To understand the impact of serialization on proxy method invocations, we introduce two more scenarios where proxies in and out of the enclave invoke methods with a serializable parameter (respectively labelled \texttt{proxy-in$\to$out+s} and \texttt{proxy-out$\to$in+s} in Figure~\ref{fig:rmi}).
The serialized parameter is a list of 16 byte string values.
We vary the size of the serialized list while keeping the number of method invocations constant at 10,000 invocations.
For scenarios \texttt{proxy-in$\to$out} and \texttt{proxy-out$\to$in}, we use the same methods as in the \texttt{...+s} variants but without passing the list as parameter.

Figure \ref{fig:rmi}\,(b) shows that RMIs in the enclave with the serialized parameter are about 10$\times$ more expensive than the corresponding RMIs without serialization, while RMIs out of the enclave are about 3$\times$ more expensive than the corresponding RMIs without serialization.

It should be noted that the orders of magnitude and ratios calculated will vary depending on the latency of the method operations themselves, without taking into account the cost of parameter serializations or enclave transitions.
The methods used in these experiments are setter methods updating an object field, which are relatively inexpensive operations.
For more expensive methods, the cost of the method operations should outweigh the cost of enclave transitions or parameter serializations, hence decreasing the importance of the latter.

\subsection{Garbage collection performance}
\label{eval-gc}

\begin{figure}[!t]
	\centering
	\includegraphics[scale=0.67]{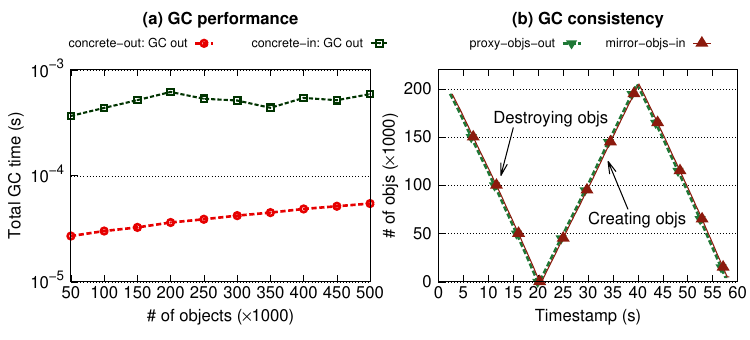}
	\caption{Garbage collection performance.}
	\label{fig:gc}
\end{figure}

\emph{(Answer to $Q_2$)}
To understand the performance variations of garbage collection in and out of the enclave, we performed an experiment which involves creating multiple concrete objects, making them eligible for GC and invoking the garbage collector in and out of the enclave.
We record the total time spent for garbage collection in both scenarios.
Figure~\ref{fig:gc}\,(a) shows the results obtained.
We observe that the enclave adds an order of magnitude more overhead to the garbage collection operation.
\graal native images embed a serial stop and copy GC~\cite{graalSite}; the copy operation of this GC in the enclave leads to more data exchange between the CPU and the EPC, hence the overhead when compared to GC performance outside (\ie, \texttt{concrete-out}). 

We performed a second experiment to demonstrate garbage collection consistency in and out of the enclave.
In this experiment a synthetic Java program creates proxy objects in the untrusted runtime, makes some of the objects eligible for GC and invokes the GC in the untrusted runtime.
This operation is repeated for a given time range.
The number of live (not garbage collected) proxy objects out of the enclave and the number of mirror objects in the enclave mirror-proxy registry are recorded at different timestamps.
Figure~\ref{fig:gc}\,(b) shows the results obtained.
We observe that as proxy objects are garbage collected, mirror objects are removed from the in-enclave mirror-proxy registry, making them eligible for GC too.
In the same way, as more proxies are created, we notice a similar increase in the number of mirror objects in the enclave.
These results show that GC is consistent between the trusted and untrusted image.

\subsection{Speed up due to partitioning}
\label{eval-perf-partitioning}

\emph{(Answer to $Q_3$)}
To demonstrate the performance improvements of partitioning native applications for enclaves with \sys, we leverage a synthetic Java program, and two real-world applications, \ie PalDB~\cite{paldb} and GraphChi~\cite{graphchi}.

\begin{figure}[!t]
	\centering
	\includegraphics[scale=0.67]{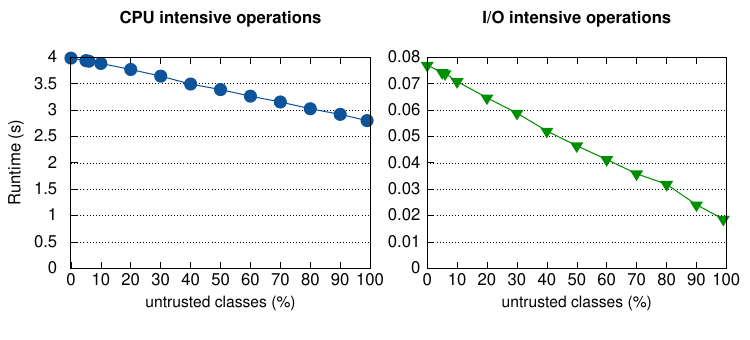}
	\caption{Enclave performance is better when fewer classes are in the enclave.}
	\label{fig:motivation}
\end{figure}

\smallskip\noindent\textbf{Synthetic benchmark.}
We developed a Java program generator to create Java applications with various numbers of classes annotated as trusted or untrusted.
We generated a Java application with 100 classes.
Each class contains an instance method which performs either CPU intensive operations (\ie, compute a fast Fourier transform~\cite{fft} on a 1\,MB double array) or I/O intensive operations (\ie, writes 4\,KB of data to a file).
The main method instantiates each class and invokes the associated instance method.
We vary the number of trusted and untrusted classes for two scenarios:
\emph{(1)}~all class instance methods perform CPU intensive operations and
\emph{(2)}~all class instance methods perform I/O intensive operations.
We then calculate the total execution time of the resulting application.
Figure~\ref{fig:motivation} shows the results.

We observe that, as the percentage of untrusted classes increases (\ie, more classes are moved out of the enclave), the overall application runtime improves.
For I/O operations, fewer enclave transitions are done for I/O write operations, leading to better performance.
For CPU operations, enclave performance can get more expensive when random reads and writes are done on data which is not present in the CPU~\cite{scone,hotcalls}.
This decrease in performance is caused by on-the-fly encryption/decryption of CPU cache-lines by the MEE~\cite{hotcalls} when data is transferred between the CPU and the EPC.
In summary, this synthetic benchmark suggests that by delegating computations to the untrusted runtime, we relieve the enclave of expensive computations, leading to better enclave performance, and better overall application performance.
We illustrate this further with the two real world applications below.

\begin{figure}[!t]
	\centering
	\includegraphics[scale=0.67]{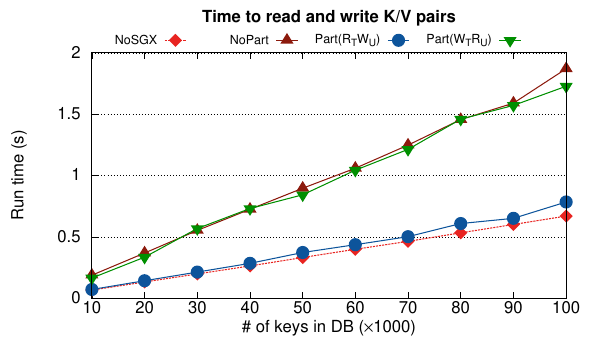}
	\caption{Read and write times for partitioned PalDB.}
	\label{fig:results-paldb}
\end{figure}

\smallskip\noindent\textbf{PalDB.}
PalDB is an embedabble persistent key-value store developed by LinkedIn, used in analytics workflows and machine-learning applications.
We consider a Java application based on PalDB which writes and reads a list of key-value (K/V) pairs in a store file.
The keys are string values of randomly generated integers (in the range $[0,2^{31}-1]$), while the values are randomly generated strings of length 128.
For this, we introduced two classes: \texttt{DBReader} and \texttt{DBWriter} which exploit PalDB's API for respectively reading from and writing to the store file.
A natural and intuitive partitioning scheme for this application is to partition along the \texttt{DBWriter} and \texttt{DBReader} classes, depending on the security requirements of the application.
For this we consider two possible scenarios: \texttt{DBReader} trusted and \texttt{DBWriter} untrusted ($R_TW_U$), and \texttt{DBReader} untrusted and \texttt{DBWriter} trusted ($R_UW_T$).
We run the unpartitioned application (base line) as a native image in an SGX enclave and compare its performance to the partitioned version with the above mentioned schemes, as well as the native image running without SGX enabled.
Figure~\ref{fig:results-paldb} shows the results obtained.

For both partitioning schemes ($R_TW_U$ and $R_UW_T$) we observe performance improvements after partitioning the application.
$R_TW_U$ is on average 2.5$\times$ times faster while $R_UW_T$ is on average 1.04$\times$ faster when compared to the unpartitioned native image.
PalDB optimises reads by memory mapping the store file in memory, but does regular I/O for writes to the store file.
This explains the greater performance improvement after partitioning using $R_TW_U$ as it relieves the enclave of expensive write-induced enclave transitions to perform I/O, making it closer to the native performance (no SGX).
For $R_UW_T$ the performance improvement is less as more \ocall transitions (23$\times$ more on average than $R_TW_U$) are done to write K/V pairs to the store file from within the enclave.
As expected, the application has best performance when running without SGX, however this is the most insecure configuration.

\begin{figure}[!t]
	\centering
	\includegraphics[scale=0.67]{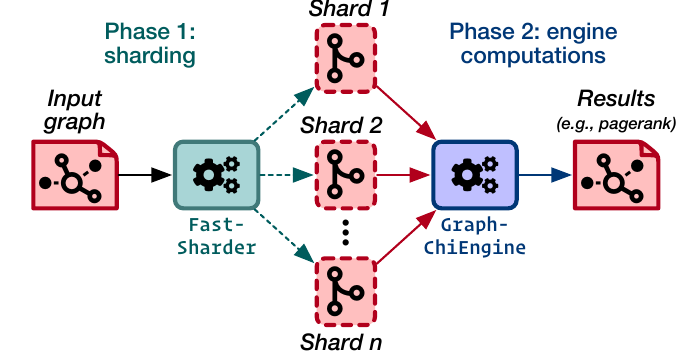}
	\caption{Typical GraphChi program workflow.}
	\label{fig:graphchi-workflow}
\end{figure}

\begin{figure}[!t]
	\centering
	\includegraphics[scale=0.67]{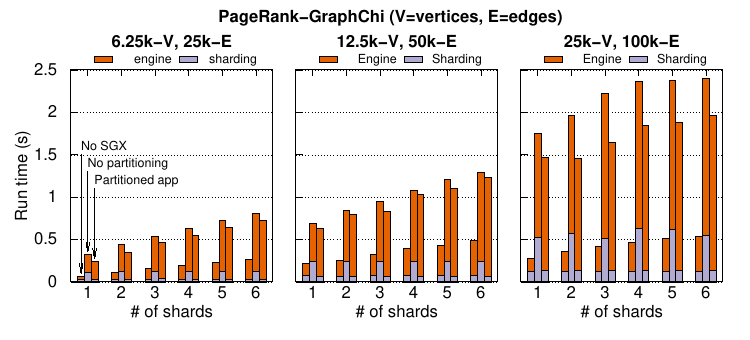}	
	\caption{Execution time for partitioned PageRank.}
	\label{fig:results-graphchi}
\end{figure}

\smallskip\noindent\textbf{GraphChi.}
GraphChi is a large-scale graph processing framework.
We use the popular \emph{PageRank}~\cite{pagerank} algorithm as an example application for partitioning.
PageRank evaluates the relative importance of nodes in a directed graph.
Typically, GraphChi applications follow the workflow outlined in Figure~\ref{fig:graphchi-workflow}.
The input graph is split by a sharder (\texttt{FastSharder}) into multiple parts (shards) which are then processed in the core execution engine (\texttt{GraphChiEngine}) to produce the final result (PageRank values in our case).
A possible partitioning scheme for the application would be along the \texttt{FastSharder} and \texttt{GraphChiEngine} classes.
For this we make the \texttt{GraphChiEngine} trusted and the \texttt{FastSharder} untrusted.
We run the PageRank algorithm on synthetic directed graphs generated using the RMAT algorithm~\cite{rmat}.
We vary graph sizes by varying the number of vertices ($V$) and edges ($E$) in the graph.
For each graph, we vary the number of shards for the PageRank computations and compare the performance of the partitioned native image to the unpartitioned case, as well as the native image running without SGX.
Figure~\ref{fig:results-graphchi} shows the results obtained.

For each shard, the leftmost bar shows the run time without SGX, the middle bar for the unpartitioned native image running inside the enclave, and the rightmost bar for the partitioned application.
We show the total times to calculate the PageRank values of the graph nodes, as well as the portion of the total time spent in sharding and in the engine. 

After partitioning, the \texttt{FastSharder} is transferred to the untrusted runtime, relieving the enclave of all expensive I/O related work done by the sharder, thereby improving enclave latency.
We can observe on the graph that the latency due to sharding after partitioning is approximately the same as the native case (no SGX), which is explained by the fact that the \texttt{FastSharder} now operates in the untrusted runtime and there is no extra overhead due to MEE encryption/decryption operations in the enclave.
This leads to a performance gain of about 1.2$\times$ on average as compared to the unpartitioned case.
We observe similar performance improvements for the different graph sizes.


\subsection{Comparing JVM-based applications in enclaves}

\label{eval-comparison-jvm}
\emph{(Answer to $Q_3$)}
To understand how partitioned and unpartitioned native images compare to the JVM-based counterparts running in enclaves, we compared SGX-based native images to the applications running on a JVM in a SCONE container.
The JVM is run with maximum heap size of 2\,GB (\texttt{-Xmx2G}).
We are not particularly concerned with the performance of applications running on a JVM out of enclaves.
However we included results for the latter to get a clearer picture of the performance variations using the different approaches.

\begin{figure}[!t]
	\centering
	\includegraphics[scale=0.67]{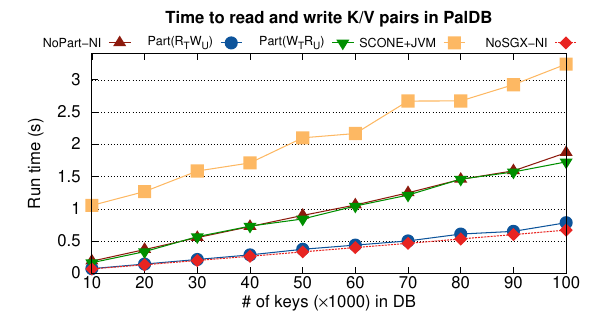}	
	\caption{Partitioned and unpartitioned PalDB native images vs. PalDB in SCONE+JVM.}
	\label{fig:scone-paldb}
\end{figure}

\begin{figure}[!t]
	\centering
	\includegraphics[scale=0.67]{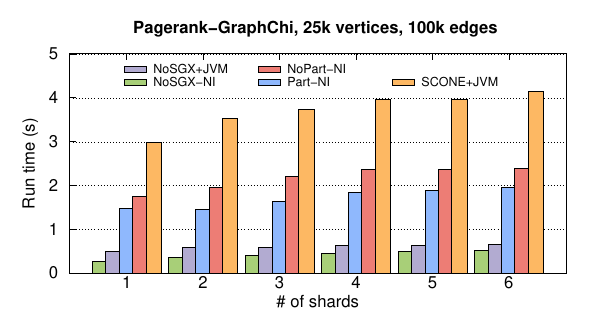}	
	\caption{Partitioned and unpartitioned GraphChi native images vs. GraphChi in SCONE+JVM.}
	\label{fig:scone-graphchi}
\end{figure}

\smallskip\noindent\textbf{Partitioned native images vs. JVM-based applications in enclaves.}
For this experiment we compared the partitioned versions of PalDB and GraphChi, using the same partitioning schemes, to their unpartitioned counterparts running on a JVM in a SCONE container.
Figures~\ref{fig:scone-paldb} and~\ref{fig:scone-graphchi} show the results for PalDB and GraphChi respectively.


From the results, we observe that $R_TW_U$ and $R_UW_T$ are respectively 6.6$\times$ and 2.8$\times$ faster on average when compared to PalDB running on a JVM in SCONE.
As for GraphChi, the partitioned GraphChi native image is 2.2$\times$ faster on average when compared to GraphChi running on a JVM in SCONE.
The poor performance of the applications with the JVM in SCONE can be justified by two reasons:
\emph{(1)}~the JVM spends some time for class loading, bytecode interpretation and dynamic compilation; these operations are absent in native images,
\emph{(2)}~the in-enclave JVM increases the number objects in the enclave heap, which leads to more data exchange between the EPC and CPU, hence more expensive MEE encryption/decryption of CPU cache lines when compared to the native images in the enclaves. 

\begin{figure}[!t]
	\centering
	\includegraphics[scale=0.67]{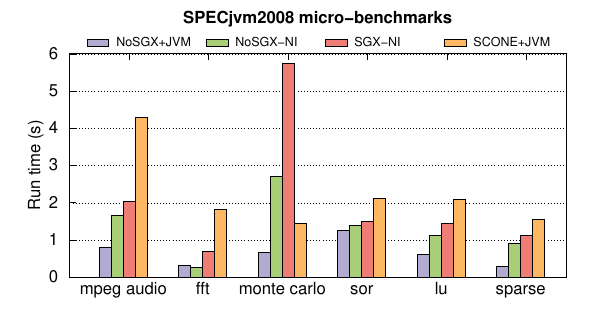}	
	\caption{Performance of unpartitioned SPECjvm 2008 micro-benchmarks in enclaves.}
	\label{fig:specjvm}
\end{figure}

\smallskip\noindent\textbf{Unpartitioned native images in enclaves vs. JVM-based applications in enclaves}.
Here we compare the performance of unpartitioned native images in enclaves to their JVM-based counterparts.
We present results for PalDB (Figure~\ref{fig:scone-paldb}) and GraphChi (Figure~\ref{fig:scone-graphchi}), as well as 6 SPECjvm2008~\cite{specjvm} micro-benchmarks (Figure~\ref{fig:specjvm}) using their default workloads.

The unpartitioned PalDB and GraphChi native images in the enclave are respectively 2.6$\times$ and 1.7$\times$ faster when compared to their JVM counterparts in a SCONE container.
For the SPECjvm2008 micro-benchmarks, Table~\ref{tab:specjvm} summarises the comparisons of the native images vs. their JVM counterparts, all running within enclaves.
We observe comparatively lower performance for the JVM counterparts in a SCONE container except for the \emph{Monte\_Carlo} micro-benchmark.
We explain this to garbage collection cycles triggered in the native image.
Recent studies~\cite{gcPerfs} suggest the GC in \graalni performs poorly when compared to OpenJDK HotSpot JVM's garbage collectors.
The poorer in-enclave JVM results for the rest can be justified by the two reasons mentioned previously.


\begin{table}[!t]		
		\setlength{\tabcolsep}{3pt}
		 \rowcolors{1}{gray!10}{gray!0}
		\begin{tabular}{l|c}
			 \rowcolor{gray!25}
			\textbf{Benchmark name} & \textbf{Latency gain over SCONE+JVM} \\
			\hline			
			Mpegaudio & $2.12\times$\\
			FFT & $2.66\times$\\
			Monte\_Carlo & $0.25\times$\\
			SOR & $1.42\times$\\
			LU & $1.46\times$\\
			Sparse & $1.38\times$\\				
			\bottomrule
		\end{tabular}
		\caption{Ratio between unpartitioned SPECjvm2008 native images in enclaves (\texttt{SGX-NI} in Figure~\ref{fig:specjvm}) against their on-JVM counterparts in SCONE (\texttt{SCONE+JVM}).}\label{tab:specjvm}
\end{table}

\subsection{Additional use-case scenarios}
\sys can be used for a wide variety of security applications. 
Examples include secure key/value stores and blockchain applications. 
The classes/business logic for storing and retrieving key/value pairs, and business logic for smart contracts can be secured in the enclave, while classes for network-related functionality are kept out of the enclave. The partitioned components then interact in accordance with our design.


\section{Conclusion and future work}
\label{sec:conclusion}

This paper presented \sys, a tool for automatically partitioning Java applications destined for secure enclave environments.
\sys leverages source code annotations and bytecode transformations to partition application classes into trusted and untrusted versions.
\sys provides an RMI-like mechanism to enable object communication between the partitioned components, as well as a garbage collection extension to ensure consistent garbage collection across the trusted and untrusted application heaps.
We implemented \sys atop \graalni, and our extensive evaluations show \sys can provide strong security guarantees while improving application performance as compared to systems that execute complete applications together with the associated runtime in an enclave.
We intend to extend this work along the following directions.
First, we will improve \sys's RMI system with transition-less cross-enclave calls for expensive RMIs, similar to~\cite{switchless}, especially useful for applications performing several enclave transitions.
Second, we plan to extend our proxy-mirror system to permit creation and interaction of proxy-mirror object pairs across multiple isolates in both the trusted and untrusted runtimes.


\begin{acks}	
	This work is supported in part by Oracle donations CR 2901 and CR 3801, as well as project 200021\_178822 of the Swiss National Science Foundation (FNS). We would also like to thank the anonymous reviewers and our shepherd, Roy H. Campbell, for their feedback.	
\end{acks}
\vspace{5mm}

{\small
\bibliographystyle{plain}
\bibliography{main.bib}}

\begin{thebibliography}{10}

\bibitem{alpine}
{Alpine Linux}.
\newblock \url{https://alpinelinux.org/}.

\bibitem{pagerank}
Alon Altman and Moshe Tennenholtz.
\newblock Ranking systems: the {PageRank} axioms.
\newblock In {\em Proceedings of the 6th {ACM} conference on {E}lectronic
  {C}ommerce (EC 05)}, Vancouver, BC, Canada, 2005.

\bibitem{trustzone}
Tiago Alves.
\newblock Trustzone: Integrated hardware and software security.
\newblock {\em White paper}, 2004.

\bibitem{scone}
Sergei Arnautov, Bohdan Trach, Franz Gregor, Thomas Knauth, Andre Martin,
  Christian Priebe, Joshua Lind, Divya Muthukumaran, Dan
  O{\textquoteright}Keeffe, Mark~L. Stillwell, David Goltzsche, Dave Eyers,
  R{\"u}diger Kapitza, Peter Pietzuch, and Christof Fetzer.
\newblock {SCONE}: Secure {Linux} containers with intel {SGX}.
\newblock In {\em 12th {USENIX} Symposium on Operating Systems Design and
  Implementation ({OSDI} 16)}, Savannah, GA, USA, 2016.

\bibitem{barnett2007annotations}
Mike Barnett, Manuel Fahndrich, Francesco Logozzo, and Diego Garbervetsky.
\newblock {Annotations for (more) Precise Points-to Analysis}.
\newblock In {\em International Workshop on Aliasing, Confinement and Ownership
  in object-oriented programming (IWACO 2007)}, Berlin, Germany, 2007.

\bibitem{saBarros}
Paulo Barros, Rene Just, Suzanne Millstein, Paul Vines, Werner Dietl, Marcelo
  d'Amorim, and Michael~D. Ernst.
\newblock Static analysis of implicit control flow: Resolving java reflection
  and android intents (t).
\newblock In {\em 2015 30th IEEE/ACM International Conference on Automated
  Software Engineering (ASE)}, pages 669--679, 2015.

\bibitem{haven}
Andrew Baumann, Marcus Peinado, and Galen Hunt.
\newblock Shielding applications from an untrusted cloud with {Haven}.
\newblock {\em ACM Transactions on Computer Systems (TOCS)}, 33(3):1--26, 2015.

\bibitem{bonettagraalvm}
Daniele Bonetta.
\newblock {GraalVM}: metaprogramming inside a polyglot system (invited talk).
\newblock In {\em Proceedings of the 3rd ACM SIGPLAN International Workshop on
  Meta-Programming Techniques and Reflection (SPLASH 18)}, Boston, MA, USA,
  2018.

\bibitem{secKeeper}
Stefan Brenner, Colin Wulf, David Goltzsche, Nico Weichbrodt, Matthias Lorenz,
  Christof Fetzer, Peter Pietzuch, and R\"{u}diger Kapitza.
\newblock {SecureKeeper}: Confidential {ZooKeeper} using intel {SGX}.
\newblock In {\em Proceedings of the 17th International Middleware Conference
  (Middleware 16)}, New York, NY, USA, 2016.

\bibitem{van2018foreshadow}
Jo~Van Bulck, Marina Minkin, Ofir Weisse, Daniel Genkin, Baris Kasikci, Frank
  Piessens, Mark Silberstein, Thomas~F. Wenisch, Yuval Yarom, and Raoul
  Strackx.
\newblock Foreshadow: Extracting the keys to the {Intel SGX} kingdom with
  transient out-of-order execution.
\newblock In {\em 27th {USENIX} Security Symposium ({USENIX} Security 18)},
  Baltimore, MD, USA, 2018.

\bibitem{rmat}
Deepayan Chakrabarti, Yiping Zhan, and Christos Faloutsos.
\newblock {R-MAT}: A recursive model for graph mining.
\newblock In {\em Proceedings of the 2004 SIAM International Conference on Data
  Mining (SDM 04)}, Lake Buena Vista, FL, USA, 2004.

\bibitem{opera}
Guoxing Chen, Yinqian Zhang, and Ten-Hwang Lai.
\newblock {OPERA}: Open remote attestation for {Intel}'s secure enclaves.
\newblock In {\em Proceedings of the 2019 ACM SIGSAC Conference on Computer and
  Communications Security (CCS 19)}, London, United Kingdom, 2019.

\bibitem{chen2021voltpillager}
Zitai Chen, Georgios Vasilakis, Kit Murdock, Edward Dean, David Oswald, and
  Flavio~D Garcia.
\newblock {VoltPillager}: Hardware-based fault injection attacks against intel
  {SGX} enclaves using the {SVID} voltage scaling interface.
\newblock In {\em 30th {USENIX} Security Symposium ({USENIX} Security 21)},
  Online, 2021.

\bibitem{fft}
W.T. Cochran, J.W. Cooley, D.L. Favin, H.D. Helms, R.A. Kaenel, W.W. Lang, G.C.
  Maling, D.E. Nelson, C.M. Rader, and P.D. Welch.
\newblock What is the fast fourier transform?
\newblock {\em Proceedings of the IEEE}, 55(10):1664--1674, 1967.

\bibitem{sgxsdk}
Intel Corporation.
\newblock {SDK for Intel\textcopyright{} Software Guard Extensions}.
\newblock
  \url{https://software.intel.com/content/www/us/en/develop/topics/software-guard-extensions/sdk.html}.

\bibitem{vcostan}
Victor Costan and Srinivas Devadas.
\newblock Intel {SGX} explained.
\newblock {\em {IACR} Cryptology ePrint Archive}, 2016(86):1--186, 2016.

\bibitem{docker}
{Docker, inc}.
\newblock {Docker: Empowering App Development for Developers}.
\newblock \url{https://www.docker.com}.

\bibitem{fegade2020scalable}
Pratik Fegade and Christian Wimmer.
\newblock Scalable pointer analysis of data structures using semantic models.
\newblock In {\em Proceedings of the 29th International Conference on Compiler
  Construction (CC 2020)}, San Diego, CA, USA, 2020.

\bibitem{hadoop}
Apache~Software Foundation.
\newblock {Apache Hadoop}.
\newblock \url{https://hadoop.apache.org/}.

\bibitem{spark}
Apache~Software Foundation.
\newblock {Apache Spark -- Unified Analytics Engine for Big Data}.
\newblock \url{https://spark.apache.org/}.

\bibitem{zookeeper}
Apache~Software Foundation.
\newblock {Apache ZooKeeper}.
\newblock \url{https://zookeeper.apache.org/}.

\bibitem{rmi}
William Grosso.
\newblock {\em Java {RMI}}.
\newblock O'Reilly Media, 2002.

\bibitem{gruss2017strong}
Daniel Gruss, Julian Lettner, Felix Schuster, Olga Ohrimenko, Istv{\'{a}}n
  Haller, and Manuel Costa.
\newblock Strong and efficient cache side-channel protection using hardware
  transactional memory.
\newblock In {\em 26th {USENIX} Security Symposium, ({USENIX} Security 2017)},
  Vancouver, BC, Canada, 2017.

\bibitem{coldboot}
J.~Alex Halderman, Seth~D. Schoen, Nadia Heninger, William Clarkson, William
  Paul, Joseph~A. Calandrino, Ariel~J. Feldman, Jacob Appelbaum, and Edward~W.
  Felten.
\newblock Lest we remember: Cold-boot attacks on encryption keys.
\newblock {\em Commun. ACM}, 52(5):91–98, May 2009.

\bibitem{jassist}
Jboss-Javassist.
\newblock {Javassist}.
\newblock \url{https://www.javassist.org/}.

\bibitem{uranus}
Jianyu Jiang, Xusheng Chen, TszOn Li, Cheng Wang, Tianxiang Shen, Shixiong
  Zhao, Heming Cui, Cho-Li Wang, and Fengwei Zhang.
\newblock Uranus: Simple, efficient {SGX} programming and its applications.
\newblock In {\em Proceedings of the 15th ACM Asia Conference on Computer and
  Communications Security (ASIACCS 2020)}, Taipei, Taiwan, 2020.

\bibitem{amd}
David Kaplan, Jeremy Powell, and Tom Woller.
\newblock {AMD} memory encryption.
\newblock {\em White paper}, 2016.

\bibitem{gcPerfs}
Jonatan Kazmierczak.
\newblock {High performance at low cost – choose the best JVM and the best
  Garbage Collector for your needs}.
\newblock \url{https://bit.ly/3f6mLjO}.

\bibitem{graphchi}
Aapo Kyrola, Guy Blelloch, and Carlos Guestrin.
\newblock Graphchi: Large-scale graph computation on just a {PC}.
\newblock In {\em 10th {USENIX} Symposium on Operating Systems Design and
  Implementation ({OSDI} 12)}, Boston, MA, USA, 2012.

\bibitem{glamdring}
Joshua Lind, Christian Priebe, Divya Muthukumaran, Dan
  O{\textquoteright}Keeffe, Pierre-Louis Aublin, Florian Kelbert, Tobias
  Reiher, David Goltzsche, David Eyers, R{\"u}diger Kapitza, Christof Fetzer,
  and Peter Pietzuch.
\newblock Glamdring: Automatic application partitioning for {Intel SGX}.
\newblock In {\em 2017 {USENIX} Annual Technical Conference ({USENIX} {ATC}
  17)}, Santa Clara, CA, USA, 2017.

\bibitem{paldb}
LinkedIn.
\newblock {PalDB}: an embeddable write-once key-value store written in {Java}.
\newblock \url{https://github.com/linkedin/PalDB}.

\bibitem{r9049642}
A.~Mogage, R.~Pires, V.~Craciun, E.~Onica, and P.~Felber.
\newblock Supply chain malware targets sgx: Take care of what you sign.
\newblock In {\em 2019 38th Symposium on Reliable Distributed Systems (SRDS)},
  pages 52--528, Los Alamitos, CA, USA, oct 2019. IEEE Computer Society.

\bibitem{oleksenko2018varys}
Oleksii Oleksenko, Bohdan Trach, Robert Krahn, Mark Silberstein, and Christof
  Fetzer.
\newblock Varys: Protecting {SGX} enclaves from practical side-channel attacks.
\newblock In {\em 2018 {USENIX} Annual Technical Conference ({USENIX} {ATC}
  2018)}, Boston, MA, USA, 2018.

\bibitem{graalFinalizers}
Oracle.
\newblock {Finalizers}.
\newblock
  \url{https://www.graalvm.org/reference-manual/native-image/Limitations/#finalizers}.

\bibitem{graalSite}
Oracle.
\newblock {GraalVM Native Image}.
\newblock \url{https://www.graalvm.org/reference-manual/native-image/}.

\bibitem{ccharpointer}
Oracle.
\newblock {GraalVM SDK Java API Reference -- CCharPointer}.
\newblock
  \url{https://www.graalvm.org/sdk/javadoc/index.html?org/graalvm/nativeimage/c/type/CCharPointer.html}.

\bibitem{centrypoint}
Oracle.
\newblock {GraalVM SDK Java API Reference -- CEntryPoint}.
\newblock
  \url{https://www.graalvm.org/sdk/javadoc/org/graalvm/nativeimage/c/function/CEntryPoint.html}.

\bibitem{javaSpecs}
Oracle.
\newblock {The Java Language Specification}.
\newblock \url{https://docs.oracle.com/javase/specs/jls/se8/jls8.pdf}.

\bibitem{agent}
Oracle.
\newblock Tracing agent.
\newblock
  \url{https://docs.oracle.com/en/graalvm/enterprise/19/guide/reference/native-image/tracing-agent.html}.

\bibitem{sgxlkl}
Christian Priebe, Divya Muthukumaran, Joshua Lind, Huanzhou Zhu, Shujie Cui,
  Vasily~A Sartakov, and Peter Pietzuch.
\newblock {SGX-LKL}: Securing the host {OS} interface for trusted execution.
\newblock {\em arXiv preprint arXiv:1908.11143}, 2019.

\bibitem{rivest1992md5}
Ronald Rivest and S~Dusse.
\newblock The md5 message-digest algorithm.

\bibitem{rountevPointsTo}
Atanas Rountev, Ana Milanova, and Barbara~G. Ryder.
\newblock Points-to analysis for {Java} using annotated constraints.
\newblock {\em SIGPLAN Not.}, 36(11):43–55, October 2001.

\bibitem{saltzer1975protection}
Jerome~H Saltzer and Michael~D Schroeder.
\newblock The protection of information in computer systems.
\newblock {\em Proceedings of the IEEE}, 63(9):1278--1308, September 1975.

\bibitem{vc3}
Felix Schuster, Manuel Costa, C{\'e}dric Fournet, Christos Gkantsidis, Marcus
  Peinado, Gloria Mainar-Ruiz, and Mark Russinovich.
\newblock {VC3}: Trustworthy data analytics in the cloud using {SGX}.
\newblock In {\em 2015 IEEE Symposium on Security and Privacy (SSP 15)}, San
  Jose, CA, USA, 2015.

\bibitem{schwarz2017malware}
Michael Schwarz, Samuel Weiser, Daniel Gruss, Cl{\'e}mentine Maurice, and
  Stefan Mangard.
\newblock {Malware guard extension: Using SGX to conceal cache attacks}.
\newblock In {\em International Conference on Detection of Intrusions and
  Malware, and Vulnerability Assessment (DIMVA 17)}, Bonn, Germany, 2017.

\bibitem{sgx-shield}
Jaebaek Seo, Byoungyoung Lee, Seong~Min Kim, Ming{-}Wei Shih, Insik Shin,
  Dongsu Han, and Taesoo Kim.
\newblock {SGX-Shield}: Enabling address space layout randomization for {SGX}
  programs.
\newblock In {\em 24th Annual Network and Distributed System Security Symposium
  (NDSS 17)}, San Diego, CA, USA, 2017.

\bibitem{panoply}
Shweta Shinde, Dat~Le Tien, Shruti Tople, and Prateek Saxena.
\newblock Panoply: Low-{TCB} linux applications with {SGX} enclaves.
\newblock In {\em 24th Annual Network and Distributed System Security Symposium
  (NDSS 17)}, San Diego, CA, USA, 2017.

\bibitem{balatsouras}
Yannis Smaragdakis and George Balatsouras.
\newblock Pointer analysis.
\newblock {\em Found. Trends Program. Lang.}, 2(1):1–69, April 2015.

\bibitem{specjvm}
SPEC.
\newblock {SPECjvm2008}.
\newblock \url{https://www.spec.org/jvm2008/}.

\bibitem{vault}
Meysam Taassori, Ali Shafiee, and Rajeev Balasubramonian.
\newblock {VAULT}: Reducing paging overheads in {SGX} with efficient integrity
  verification structures.
\newblock In {\em Proceedings of the Twenty-Third International Conference on
  Architectural Support for Programming Languages and Operating Systems (ASPLOS
  16)}, ASPLOS '18, page 665–678, Williamsburg, VA, USA, 2018.

\bibitem{switchless}
Hongliang Tian, Qiong Zhang, Shoumeng Yan, Alex Rudnitsky, Liron Shacham, Ron
  Yariv, and Noam Milshten.
\newblock Switchless calls made practical in intel {SGX}.
\newblock In {\em Proceedings of the 3rd Workshop on System Software for
  Trusted Execution ({SysTEX} 18)}, New York, NY, USA, 2018.

\bibitem{graphene}
Chia-Che Tsai, Donald~E. Porter, and Mona Vij.
\newblock {Graphene-SGX}: {A} practical library {OS} for unmodified
  applications on {SGX}.
\newblock In {\em 2017 {USENIX} Annual Technical Conference ({USENIX} {ATC}
  2017)}, Santa Clara, CA, USA, 2017.

\bibitem{civet}
Chia-Che Tsai, Jeongseok Son, Bhushan Jain, John McAvey, Raluca~Ada Popa, and
  Donald~E. Porter.
\newblock Civet: An efficient {Java} partitioning framework for hardware
  enclaves.
\newblock In {\em 29th {USENIX} Security Symposium ({USENIX} Security 20)},
  Online, 2020.

\bibitem{cfhider}
Yongzhi Wang, Yulong Shen, Cuicui Su, Ke~Cheng, Yibo Yang, Anter Faree, and Yao
  Liu.
\newblock {CFHider}: Control flow obfuscation with {Intel SGX}.
\newblock In {\em {IEEE Conference on Computer Communications (INFOCOM 2019)}},
  Paris, France, 2019.

\bibitem{sgxperf}
Nico Weichbrodt, Pierre{-}Louis Aublin, and R{\"{u}}diger Kapitza.
\newblock sgx-perf: {A} performance analysis tool for intel {SGX} enclaves.
\newblock In Paulo Ferreira and Liuba Shrira, editors, {\em Proceedings of the
  19th International Middleware Conference (Middleware 2018)}, Rennes, France,
  2018.

\bibitem{hotcalls}
Ofir Weisse, Valeria Bertacco, and Todd Austin.
\newblock Regaining lost cycles with hotcalls: A fast interface for sgx secure
  enclaves.
\newblock {\em SIGARCH Comput. Archit. News}, 45(2):81–93, June 2017.

\bibitem{nativeImgs}
Christian Wimmer, Codrut Stancu, Peter Hofer, Vojin Jovanovic, Paul
  W{\"o}gerer, Peter~B Kessler, Oleg Pliss, and Thomas W{\"u}rthinger.
\newblock Initialize once, start fast: application initialization at build
  time.
\newblock In {\em 2019 ACM SIGPLAN International Conference on Object-Oriented
  Programming, Systems, Languages and Applications (OOPSLA 19)}. Athens,
  Greece, 2019.

\bibitem{busprobing}
Zhenyu Xu, Thomas Mauldin, Qing Yang, and Tao Wei.
\newblock Runtime detection of probing/tampering on interconnecting buses.
\newblock In {\em 2021 IEEE 29th Annual International Symposium on
  Field-Programmable Custom Computing Machines (FCCM)}, pages 247--251, 2021.

\bibitem{yuhala2021plinius}
P.~Yuhala, P.~Felber, V.~Schiavoni, and A.~Tchana.
\newblock Plinius: Secure and persistent machine learning model training.
\newblock In {\em 2021 51st Annual IEEE/IFIP International Conference on
  Dependable Systems and Networks (DSN)}, pages 52--62, Los Alamitos, CA, USA,
  jun 2021. IEEE Computer Society.

\bibitem{opaque}
Wenting Zheng, Ankur Dave, Jethro~G Beekman, Raluca~Ada Popa, Joseph~E
  Gonzalez, and Ion Stoica.
\newblock Opaque: An oblivious and encrypted distributed analytics platform.
\newblock In {\em 14th {USENIX} Symposium on Networked Systems Design and
  Implementation (NSDI 17)}, Boston, MA, USA, 2017.

\end{thebibliography}

\end{document}